\documentclass[11pt,a4paper]{article}

\usepackage{epsfig,amsmath,amssymb,cite}
\usepackage{hyperref}
\usepackage{graphicx}

\begin{document}

\title{Thin shells in (2+1)-dimensional $F(R)$ gravity} 
\author{Ernesto F. Eiroa\thanks{e-mail: eiroa@iafe.uba.ar} , Griselda Figueroa-Aguirre\thanks{e-mail: gfigueroa@iafe.uba.ar}\\
{\small  Instituto de Astronom\'{\i}a y F\'{\i}sica del Espacio (IAFE, CONICET-UBA),}\\
{\small Casilla de Correo 67, Sucursal 28, 1428, Buenos Aires, Argentina}} 
\date{}
\maketitle

\begin{abstract}
We study thin shells of matter in (2+1)-dimensional $F(R)$ theories of gravity with constant scalar curvature $R$. We consider a wide class of spacetimes with circular symmetry, in which a thin shell joins an inner region with an outer one. We analyze the stability of the static configurations under radial perturbations. As examples of spacetimes asymptotically anti--de Sitter, we present a charged bubble and a charged thin shell surrounding a non-charged black hole. In both cases, we show that stable solutions can be found for suitable values of the parameters.
\end{abstract}

\section{Introduction}\label{intro} 

The astronomical observations pose challenges to theoretical physics that have not yet been fully resolved. The accelerated expansion of the Universe during the matter dominated epoch needs, within the framework of general relativity, the presence of dark energy, while the structure formation requires the presence of dark matter. The explanation of the observed cosmic microwave background also demands them, besides the ordinary matter and the electromagnetic radiation. Modified gravity theories were proposed in order to solve both problems without the need for non-standard fluids or a fine tuning of the cosmological constant, required by the concordance ($\Lambda$CDM) model. Among them are the well known $F(R)$ theories \cite{revfr}, in which the Einstein--Hilbert Lagrangian is replaced by a function $F(R)$ of the Ricci scalar curvature $R$. The adoption of $F(R)$ gravity can provide an alternative for a unified picture of both inflation and the accelerated expansion at later times. Besides the cosmological aspects,  the study of black holes \cite{bhfr,bhrnonconst}, branes \cite{branefr}, and traversable wormholes \cite{whfr}, has been of interest in $F(R)$ theories.

The junction conditions in general relativity \cite{daris} allow for a proper matching of two solutions through a hypersurface. They demand the continuity of the first fundamental form there and the result is a boundary hypersurface or a thin shell of matter. In the last case, the characteristics and the dynamics of the matter at the matching hypersurface can be analyzed by using this formalism. It is a useful tool in order to develop models of thin layers of matter --that may surround vacuum (bubbles) or black holes-- \cite{sh1,sh2}, wormholes \cite{wh01,wh02,whcil}, and gravastars \cite{gravstar}, among others. In highly symmetric scenarios, e.g. spherical or cylindrical, the stability analysis of the configurations can be performed rather easily. In recent years, the junction formalism has been extended to $F(R)$ theories \cite{dss,js1}. Unlike general relativity, $F(R)$ gravity additionally requires the continuity of the trace of the second fundamental form at the joining hypersurface and, in the case of non-quadratic $F(R)$, also the continuity of the scalar curvature there \cite{js1}. For quadratic $F(R)$, this second condition can be relaxed, and new contributions to the standard energy-momentum tensor are incorporated, that is, an external scalar pressure/tension, an external energy flux vector, and a double layer energy-momentum tensor which resembles classical dipole distributions \cite{js1,js2}. The junction formalism has subsequently been generalized to the most generic gravitational theory with a Lagrangian containing terms quadratic in the curvature \cite{rsv,bdes}, which shares the main features of quadratic $F(R)$. In the general quadratic scenario, the existence of a double layer can be fully derived by using distributions \cite{rsv} or only from the least action principle \cite{bdes}; in this last case, the presence of the derivative of the Dirac delta function is avoided in the calculation and the Dirac delta functions are cancelled before integration. The junction conditions in $F(R)$ gravity have been adopted to model bubbles \cite{eftsfr,efncrfr}, thin shells of matter surrounding black holes \cite{eftsfr,efncrfr}, and thin-shell wormholes \cite{efwhfr,whtsfr}. The interesting case of pure double layers have also been addressed \cite{pdlfr}.

Low dimensional spacetimes are of interest because they can help to address some conceptual problems related to black hole physics, quantum gravity, and string theory \cite{carlip}.  In (2+1) general relativity, black holes \cite{btz}, polytropic stars \cite{ps3d}, wormholes \cite{wh3d}, gravastars \cite{grava3d}, thin shells of matter \cite{ts3d}, and thin-shell wormholes \cite{whts3d} have been research topics. Black hole solutions in (2+1)-dimensional $F(R)$ gravity have been found in recent years \cite{hendi14a,hendi14b,hendi20}. But low dimensional thin shells in these theories have not been much explored. In this paper, we use the junction conditions for spacetimes with (2+1) dimensions to construct circular thin shells of matter in $F(R)$ gravity with constant scalar curvature $R$. The global topology consists of an inner geometry joined across the shell to an outer one. We analyze the stability of the static configurations under perturbations that preserve the symmetry. We consider two examples of anti--de Sitter spacetimes with a conformally invariant Maxwell field as a source: a charged bubble and a charged thin shell of matter enclosing a non-charged black hole. In Sec. \ref{jc} we review the junction conditions for proper matching, in Sec. \ref{tsh} we introduce the general   procedure for the construction of circular shells, while in Sec. \ref{examples} we show the different examples. Finally, in Sec. \ref{conclu} we discuss the results obtained. We adopt units so that $c=G=1$, with $c$ the speed of light and $G$ the gravitational constant.

\section{Junction conditions in (2+1)-dimensional $F(R)$ theories}\label{jc}

We start by reviewing the (2+1)-dimensional junction formalism, in which a manifold $\mathcal M$ is the union of two parts $\mathcal M_1$ and $\mathcal M_2$ through a one dimensional hypersurface $\Sigma$, corresponding to a boundary hypersurface or a thin shell, depending on the particular situation. We denote the first fundamental form on $\Sigma$ by $h^{1,2}_{\mu \nu}$ and the second fundamental form (or extrinsic curvature) by $K^{1,2}_{\mu \nu}$, where the superscripts refer to the corresponding parts of $\mathcal M$. The jump of any quantity $\Upsilon $ across $\Sigma$ is defined by $[\Upsilon ]\equiv (\Upsilon ^{2}-\Upsilon ^{1})|_\Sigma$. In $F(R)$ gravity one has to demand \cite{js1}, as in general relativity, the continuity at $\Sigma$ of the first fundamental form inherited from both $\mathcal M_{1,2}$, which gives
\begin{equation}
[h_{\mu \nu}]=0. 
\label{fffjump}
\end{equation}
As a consequence, the unit normal $n^\mu$ to $\Sigma$ is well defined without a jump at $\Sigma$, but for computational purposes the expressions at both sides of $\Sigma$ are usually given. In $F(R)$ theories, there is the additional condition \cite{js1} that the trace of the second fundamental form has a null jump at $\Sigma$
\begin{equation}
[K^{\mu}_{\;\; \mu}]=0 
\label{traceKjump}
\end{equation}
and, when  $F'''(R) \neq 0$ (the prime means the derivative with respect to $R$), the continuity of $R$ across $\Sigma$ is also required \cite{js1}, i.e. $[R]=0$. In this case, the energy-momentum tensor $S_{\mu \nu}$ at the joining hypersurface $\Sigma$ has the form
\begin{equation}
\kappa  S_{\mu \nu}=-F'(R_\Sigma)[K_{\mu \nu}]+ F''(R_\Sigma)[\eta^\gamma \nabla_\gamma R]  h_{\mu \nu}, \;\;\;\; n^{\mu}S_{\mu\nu}=0,
\label{LanczosGen}
\end{equation}
where $\kappa =8\pi $ and $\nabla$ is the covariant derivative. There is an ordinary boundary hypersurface at $\Sigma$ when $S_{\mu \nu}=0$ or a thin shell of matter otherwise. However, when $F'''(R)= 0$, that is $F(R)$ quadratic theory 
\begin{equation}
F(R)=R -2\Lambda +\alpha R^2 
\label{quadF}
\end{equation}
where $\alpha$ is a parameter and $\Lambda$ is the cosmological constant, a discontinuity of $R$ at $\Sigma$ is allowed \cite{js1,js2}. The energy-momentum tensor at $\Sigma$ in the quadratic case takes the form \cite{js1,js2,rsv}
\begin{equation}
\kappa S_{\mu \nu} =-[K_{\mu\nu}]+2\alpha( [n^{\gamma }\nabla_{\gamma}R] h_{\mu\nu}-[RK_{\mu\nu}]),  \qquad  n^{\mu}S_{\mu\nu}=0;
\label{LanczosQuad}
\end{equation}
besides this one, there are also three other contributions \cite{js1,js2,rsv}; namely, an external energy flux vector
\begin{equation}
\kappa\mathcal{T}_\mu=-2\alpha \overline{\nabla}_\mu[R],  \qquad  n^{\mu}\mathcal{T}_\mu=0,
\label{Tmu}
\end{equation}
where $\overline{\nabla }$ is the intrinsic covariant derivative on $(\Sigma,h_{\mu\nu})$; an external scalar pressure or tension
\begin{equation}
\kappa\mathcal{T}=2\alpha [R] K^\gamma{}_\gamma ;
\label{Tg}
\end{equation}
and a two-covariant symmetric tensor distribution 
\begin{equation}
\kappa \mathcal{T}_{\mu \nu}=\nabla_{\gamma } \left( 2\alpha [R] h_{\mu \nu } n^{\gamma } \delta ^{\Sigma }\right),
\label{dualay1}
\end{equation}
where $\delta ^{\Sigma }$ denotes the Dirac delta with support on $\Sigma $, or in the equivalent form\footnote{Note that in this expression the indices of $\Psi^{\mu\nu}$ are missing in Refs. \cite{js1,js2}.}
\begin{equation}
\kappa \left<\mathcal{T}_{\mu \nu},\Psi ^{\mu \nu } \right> = -\int_\Sigma 2\alpha[R] h_{\mu \nu }  n^\gamma\nabla_\gamma \Psi ^{\mu \nu },
\label{dualay2}
\end{equation}
for any test tensor field $\Psi ^{\mu \nu }$. This double layer energy-momentum distribution $\mathcal{T}_{\mu \nu }$ corresponds to a Dirac ``delta prime'' type contribution with strength \cite{js1,js2,rsv}
\begin{equation}
\kappa \mathcal{P}_{\mu \nu } =2\alpha[R] h_{\mu \nu }, \hspace{1cm}  \mathcal{P}_{\mu \nu } = \mathcal{P}_{\nu \mu }, \hspace{1cm} n^{\mu } \mathcal{P}_{\nu \mu } =0, 
\label{strength}
\end{equation}
resembling dipole distributions in classical electrodynamics \cite{js1,js2,rsv}. All the above contributions are required in order to make the complete energy-momentum tensor divergence free \cite{js1,js2,rsv}, necessary for local conservation. If $K_{\mu \nu}$ and $R$ have no jumps at $\Sigma$, all these contributions vanish and $\Sigma $ is an ordinary boundary hypersurface. In general, in quadratic $F(R)$ there is a thin shell plus a double layer at the matching hypersurface.

\section{Circular thin shells: construction and stability}\label{tsh}

In what follows, we adopt a constant value for the curvature scalar at each side of the hypersurface $\Sigma$. We start with the circularly symmetric geometries
\begin{equation}
ds^2=-A_{1,2} (r) dt_{1,2}^2+A_{1,2} (r)^{-1} dr^2+r^2d\theta^2 ,
\label{metric-sphe}
\end{equation}
where $t_{1,2}$ is the corresponding time coordinate, $r>0$ is the radial coordinate, and $0\le \theta \le 2\pi$ is the angular coordinate. We take a radius $a$ in order to define the hypersurface (circle) $\Sigma$ by $r=a$, the inner region $\mathcal{M}_1$ by $0\leq r \leq a$, and the outer region $\mathcal{M}_2$ by $r \geq a$. We join these regions at $\Sigma$, obtaining a new manifold $\mathcal{M}=\mathcal{M}_1 \cup \mathcal{M}_2$, where a global radial coordinate  $r\in [0,+\infty)$ is defined and the angular coordinates are mutually identified. The complete spacetime $\mathcal{M}$ is described by the coordinates $X^{\alpha }_{1,2} = (t_{1,2},r,\theta)$, while on $\Sigma$ we take the coordinates $\xi ^{i}=(\tau ,\theta )$, with $\tau $ the proper time. We let the radius $a$ depend on $\tau $, i.e. $a(\tau)$, denoting its derivative with respect to $\tau$ by $\dot{a}(\tau)$. The proper time should be the same at both sides of $\Sigma$, so we have
\begin{equation*}
\frac{dt_{1,2}}{d\tau} = \frac{\sqrt{A_{1,2}(a) + \dot{a} ^2}}{A_{1,2}(a)},
\end{equation*}
in which the free signs are fixed by demanding that the times $t_{1,2}$ and $\tau$ all run into the future. The first fundamental form at the sides of the shell is given by
\begin{equation}
h^{1,2}_{ij}= \left. g^{1,2}_{\mu\nu}\frac{\partial X^{\mu}_{1,2}}{\partial\xi^{i}}\frac{\partial X^{\nu}_{1,2}}{\partial\xi^{j}}\right| _{\Sigma },
\end{equation}
and the second fundamental form reads
\begin{equation}
K_{ij}^{1,2 }=-n_{\gamma }^{1,2 }\left. \left( \frac{\partial ^{2}X^{\gamma
}_{1,2} } {\partial \xi ^{i}\partial \xi ^{j}}+\Gamma _{\alpha \beta }^{\gamma }
\frac{ \partial X^{\alpha }_{1,2}}{\partial \xi ^{i}}\frac{\partial X^{\beta }_{1,2}}{
\partial \xi ^{j}}\right) \right| _{\Sigma },
\label{sff}
\end{equation}
where the unit normals ($n^{\gamma }n_{\gamma }=1$) are determined by
\begin{equation}
\label{general_normal_fr}
n_{\gamma }^{1,2 }=\left\{ \left. \left| g^{\alpha \beta }_{1,2}\frac{\partial G}{\partial
X^{\alpha }_{1,2}}\frac{\partial G}{\partial X^{\beta }_{1,2}}\right| ^{-1/2}
\frac{\partial G}{\partial X^{\gamma }_{1,2}} \right\} \right| _{\Sigma },
\end{equation}
with $G(r)\equiv r-a =0$ at $\Sigma$, and they are taken to point from $\mathcal{M}_1$ to $\mathcal{M}_2$. On the surface $\Sigma $, we prefer to work in the orthonormal basis $\{ e_{\hat{\tau}}=e_{\tau }, e_{\hat{\theta}}=a^{-1}e_{\theta }\} $, for an easier interpretation of the results. Therefore, and taking into account the metrics (\ref{metric-sphe}), we obtain that the first fundamental form is $h^{1,2}_{\hat{\imath}\hat{\jmath}}= \mathrm{diag}(-1,1)$, the unit normals result
\begin{equation}
n_{\gamma }^{1,2}= \left(-\dot{a},\frac{\sqrt{A_{1,2}(a)+\dot{a}^2}}{A_{1,2}(a)},0 \right),
\end{equation}
while the non-null components of the second fundamental form are
\begin{equation}
K_{\hat{\theta}\hat{\theta}}^{1,2}=\frac{1}{a}\sqrt{A_{1,2} (a) +\dot{a}^2}
\end{equation}
and
\begin{equation}
K_{\hat{\tau}\hat{\tau}}^{1,2 }=-\frac{A '_{1,2}(a)+2\ddot{a}}{2\sqrt{A_{1,2}(a)+\dot{a}^2}},
\end{equation}
so their jumps at $\Sigma$ read
\begin{equation}
[K_{\hat{\tau}\hat{\tau}}]=-\frac{A '_{2}(a)+2\ddot{a}}{2\sqrt{A_{2}(a)+\dot{a}^2}}+\frac{A '_{1}(a)+2\ddot{a}}{2\sqrt{A_{1}(a)+\dot{a}^2}}
\end{equation}
and
\begin{equation}
[K_{\hat{\theta}\hat{\theta}}]=\frac{1}{a}\sqrt{A_{2} (a) +\dot{a}^2}-\frac{1}{a}\sqrt{A_{1} (a) +\dot{a}^2}.
\end{equation}
The proper matching at $\Sigma$ always requires the fulfillment of Eq. (\ref{traceKjump}), which gives
\begin{equation}
\frac{2\ddot{a}+ A_{2}'(a)}{2\sqrt{A_{2}(a)+\dot{a}^2}}-\frac{2\ddot{a}+ A_{1}'(a)}{2\sqrt{A_{1}(a)+\dot{a}^2}}+\frac{1}{a}\left(\sqrt{A_{2}(a)+\dot{a}^2}-\sqrt{A_{1}(a)+\dot{a}^2}\right)=0.
\label{cond-dyna}
\end{equation}
As previously mentioned, we have two cases to consider for the analysis of the matter content at the hypersurface $\Sigma$. In both of them, since the curvature scalar is constant in each of the regions $\mathcal{M}_1$ and $\mathcal{M}_2$, it is clear that $[\eta^\gamma \nabla_\gamma R]=0$, which simplifies Eqs. (\ref{LanczosGen}) and (\ref{LanczosQuad}). The denominated \cite{js1} brane tension $\lambda = F''(R_\Sigma) [\eta^\gamma \nabla_\gamma R]$ vanishes in this case. In the orthonormal basis, the energy-momentum tensor takes the form  $S_{_{\hat{\imath}\hat{\jmath} }}={\rm diag}(\sigma ,p)$, with $\sigma$ the energy density and $p=p_\theta$ the transverse pressure.

\subsection{Case $[R]=0$}

When $R_1=R_2= R_0$, so that $[R]=0$, for any $F(R)$ theory we find, from Eqs. (\ref{LanczosGen}) or (\ref{LanczosQuad}) as appropriate, the expressions of the energy density
\begin{equation}
\sigma= -\frac{F'(R_0)}{\kappa }\left( -\frac{2\ddot{a}+A_{2}'(a)}{2\sqrt{A_{2}(a)+\dot{a}^2}}+\frac{2\ddot{a}+A_{1}'(a)}{2\sqrt{A_{1}(a)+\dot{a}^2}}\right),
\label{energy-dyna1}
\end{equation}
which by using Eq. (\ref{cond-dyna}) can be rewritten in the form 
\begin{equation}
\sigma= -\frac{F'(R_0)}{\kappa a}\left( \sqrt{A_{2}(a)+\dot{a}^2}-\sqrt{A_{1}(a)+\dot{a}^2}\right),
\label{energy-dyna2}
\end{equation}
and the pressure
\begin{equation}
p=-\frac{F'(R_0)}{\kappa a}\left( \sqrt{A_{2}(a)+\dot{a}^2}-\sqrt{A_{1}(a)+\dot{a}^2}\right). 
\label{press-dyna1}
\end{equation}
In quadratic gravity, $F'(R_0)=1+2\alpha R_0$, while the other three contributions $\mathcal{T}$, $\mathcal{T}_\mu$, and $\mathcal{T}_{\mu\nu}$, proportional to $[R]$, are all null. It is straightforward to see that $ \sigma -p=0 $, therefore $p = \sigma$, i.e. stiff matter. For the static shell, with a radius $a_0$ satisfying the static version of Eq. (\ref{cond-dyna}), that is
\begin{equation}
\frac{ A_{2}'(a_0)}{2\sqrt{A_{2}(a_0)}}-\frac{ A_{1}'(a_0)}{2\sqrt{A_{1}(a_0)}}+\frac{1}{a_0}\left(\sqrt{A_{2}(a_0)}-\sqrt{A_{1}(a_0)}\right)=0,
\label{cond-stat}
\end{equation}
from Eqs. (\ref{energy-dyna1}), (\ref{energy-dyna2}), and (\ref{press-dyna1}), we have 
\begin{equation}
\sigma _0= -\frac{F'(R_0)}{\kappa }\left( -\frac{A_{2}'(a_0)}{2\sqrt{A_{2}(a_0)}}+\frac{A_{1}'(a_0)}{2\sqrt{A_{1}(a_0)}}\right)
\label{energy-stat1},
\end{equation}
or
\begin{equation}
\sigma _0= -\frac{F'(R_0)}{\kappa a_0} \left( \sqrt{A_{2}(a_0)}-\sqrt{A_{1}(a_0)}\right) ,
\label{energy-stat2}
\end{equation}
and
\begin{equation}
p_0=-\frac{F'(R_0)}{\kappa a_0}\left( \sqrt{A_{2}(a_0)}-\sqrt{A_{1}(a_0)}\right)
\label{press-stat1},
\end{equation}
which satisfy $\sigma _0 -p_0=0 $.

\subsection{Case $[R]\neq 0$}

When $R_1 \neq R_2$, that is $[R]\neq 0$, we should work in quadratic $F(R)$, so we have $F(R_{1,2})=R_{1,2}-2\Lambda+\alpha R_{1,2}^2$ at the sides of $\Sigma$, from which we can easily verify that $F'(R_{1,2})= 1+2\alpha R_{1,2}$ and $F''(R_{1,2})=2\alpha$. For a shell radius satisfying Eq. (\ref{cond-dyna}), from Eq. (\ref{LanczosQuad}) we obtain the energy density
\begin{equation}
\sigma= \frac{1+2\alpha R_2}{\kappa }\left( \frac{2\ddot{a}+A_{2}'(a)}{2\sqrt{A_{2}(a)+\dot{a}^2}} \right)- \frac{1+2\alpha R_1}{\kappa }\left( \frac{2\ddot{a}+A_{1}'(a)}{2\sqrt{A_{1}(a)+\dot{a}^2}}\right)
\label{energy-dyna-3}
\end{equation}
and the pressure
\begin{equation}
p= -\frac{1+2\alpha R_2}{\kappa }\left(\frac{\sqrt{A_{2}(a)+\dot{a}^2}}{a}\right)+ \frac{1+2\alpha R_1}{\kappa }\left(\frac{\sqrt{A_{1}(a)+\dot{a}^2}}{a}\right), 
\label{press-dyna-2}
\end{equation}
while, from Eq. (\ref{Tg}) and using  Eq. (\ref{cond-dyna}), the external scalar pressure or tension reads
\begin{equation}
\mathcal{T}=\frac{2\alpha R_2}{\kappa }\left( \frac{2\ddot{a}+A_{2}'(a)}{2\sqrt{A_{2}(a)+\dot{a}^2}} + \frac{\sqrt{A_{2}(a)+\dot{a}^2}}{a} \right)- \frac{2\alpha R_1}{\kappa }\left( \frac{2\ddot{a}+A_{1}'(a)}{2\sqrt{A_{1}(a)+\dot{a}^2}}+ \frac{\sqrt{A_{1}(a)+\dot{a}^2}}{a}\right);
\label{extpress-dyna}
\end{equation}
it is clear that they satisfy $\sigma -p=\mathcal{T}$. The other extra contributions at $\Sigma$ are the external energy flux vector, which from Eq. (\ref{Tmu}) results
\begin{equation}
\mathcal{T}_\mu =0 
\label{Tmu-dyna}
\end{equation}
and the double layer energy-momentum distribution $\mathcal{T}_{\mu \nu }$  with a strength that, from Eq. (\ref{strength}), in the orthonormal basis takes the form
\begin{equation}
\mathcal{P}_{\hat{\imath}\hat{\jmath}}= \frac{2 \alpha}{\kappa} [R]h_{\hat{\imath}\hat{\jmath}}.
\label{strength-dyna}
\end{equation}
In the static case, with a radius $a_0$ that should satisfy Eq. (\ref{cond-stat}), we obtain 
\begin{equation}
\sigma_0= \frac{1+2\alpha R_2}{\kappa }\left( \frac{A_{2}'(a_0)}{2\sqrt{A_{2}(a_0)}} \right)- \frac{1+2\alpha R_1}{\kappa }\left( \frac{A_{1}'(a_0)}{2\sqrt{A_{1}(a_0)}}\right),
\label{energy-stat3}
\end{equation}
\begin{equation}
p_0= -\frac{1+2\alpha R_2}{\kappa }\left(\frac{\sqrt{A_{2}(a_0)}}{a_0}\right)+ \frac{1+2\alpha R_1}{\kappa }\left(\frac{\sqrt{A_{1}(a_0)}}{a_0}\right)
\label{press-stat2},
\end{equation}
and
\begin{equation}
\mathcal{T}_0=\frac{2\alpha R_2}{\kappa }\left( \frac{A_{2}'(a_0)}{2\sqrt{A_{2}(a_0)}} + \frac{\sqrt{A_{2}(a_0)}}{a_0} \right)- \frac{2\alpha R_1}{\kappa }\left( \frac{A_{1}'(a_0)}{2\sqrt{A_{1}(a_0)}}+ \frac{\sqrt{A_{1}(a_0)}}{a_0}\right);
\label{extpress-stat}
\end{equation}
which fulfill $\sigma_0 -p_0=\mathcal{T}_0 $. The other extra contributions do not change in the static case, so that $\mathcal{T}_\mu ^{(0)}=\mathcal{T}_\mu $ and $\mathcal{P}_{\hat{\imath}\hat{\jmath}}^{(0)}= \mathcal{P}_{\hat{\imath}\hat{\jmath}} $, given by Eqs. (\ref{Tmu-dyna}) and (\ref{strength-dyna}), respectively.

\subsection{Stability}

Now, we proceed with the stability analysis of the static solutions under radial perturbations. In both cases, by using that $\ddot{a}= (1/2)d(\dot{a}^2)/da$ and defining the variable $z=\sqrt{A_{2}(a)+\dot{a}^2}-\sqrt{A_{1}(a)+\dot{a}^2}$, we rewrite Eq. (\ref{cond-dyna}) to obtain the equivalent equation $az'(a)+z(a)=0$. By solving this differential equation we find an expression for $\dot{a}^{2}$ in terms of an effective potential
\begin{equation}
\dot{a}^{2}=-V(a),
\label{condicionPot}
\end{equation}
where 
\begin{equation}
V(a)=-\frac{a_0^2 \left(\sqrt{A_2\left(a_0\right)}-\sqrt{A_1\left(a_0\right)}\right)^2}{4 a^2}-\frac{a^2 \left(A_2(a)-A_1(a)\right)^2}{4 a_0^2 \left(\sqrt{A_2\left(a_0\right)}-\sqrt{A_1\left(a_0\right)}\right)^2}+ \frac{A_1(a)+A_2(a)}{2}.
\label{pot}
\end{equation}
It is easy to verify that $V(a_0)=0$, $V'(a_0)=0$, and
\begin{eqnarray}
V''(a_0)&=& -\frac{3  \left(\sqrt{A_2\left(a_0\right)}-\sqrt{A_1\left(a_0\right)}\right)^2}{2 a_0^2}-\frac{\left(\sqrt{A_1\left(a_0\right)}+\sqrt{A_2\left(a_0\right)}\right)^2}{2 a_0^2} \nonumber  \\
&&-\frac{\left(A_2'(a_0)-A_1'(a_0)\right)^2}{2 \left(\sqrt{A_2\left(a_0\right)}-\sqrt{A_1\left(a_0\right)}\right)^2}-\frac{2 \left(A_2(a_0)-A_1(a_0)\right) \left(A_2'(a_0)-A_1'(a_0)\right)}{a_0 \left(\sqrt{A_2\left(a_0\right)}-\sqrt{A_1\left(a_0\right)}\right)^2} \nonumber \\
&&-\frac{\left(A_2(a_0)-A_1(a_0)\right) \left(A_2''(a_0)-A_1''(a_0)\right)}{2 \left(\sqrt{A_2\left(a_0\right)}-\sqrt{A_1\left(a_0\right)}\right)^2}+\frac{A_1''(a_0)+A_2''(a_0)}{2}.
\label{potd2}
\end{eqnarray}
As usual, a configuration is stable under radial perturbations when $V''(a_0)>0$.

\section{Examples of anti--de Sitter spacetimes}\label{examples}

The (2+1)-dimensional action in which the gravitational Lagrangian $F(R)=R+ f(R)$ is coupled to non-linear electrodynamics can be written in the form
\begin{equation}
I =\frac{1}{16\pi }\int d^{3}x\sqrt{-g}\left(R+ f(R) +\left( - F_{\alpha\beta}F^{\alpha\beta} \right) ^{s}\right) ,
\label{Action}
\end{equation}
where $F_{\alpha\beta}F^{\alpha\beta}$ is the Maxwell invariant, with $F_{\mu \nu }=\partial _{\mu }A_{\nu }-\partial _{\nu }A_{\mu }$ the electromagnetic tensor field and $A_{\mu }$ the gauge potential, and $s$ is an arbitrary positive non-linearity parameter ($s\neq 1/2$). The associated energy-momentum tensor is $T_{\mu\nu}=2\left[ -sF_{\mu \gamma}F_{\nu}^{\;\;\gamma}(-F_{\alpha\beta}F^{\alpha\beta})^{s-1}-(1/4) g_{\mu\nu}(-F_{\alpha\beta}F^{\alpha\beta})^s\right]$. The special case $s = 3/4$, which leads to a traceless $T_{\mu\nu}$, corresponds to the conformally invariant Maxwell field as a source \cite{hendi14a}. In this case, the field equations in the metric formalism are
\begin{equation}
R_{\mu\nu}(1+f'(R)) - \frac{1}{2}g_{\mu\nu}(R+f(R))+
(g_{\mu\nu}\nabla_\gamma \nabla^\gamma -\nabla_\mu \nabla_\nu)f'(R)=8\pi T_{\mu\nu},
\label{field_eqns}
\end{equation}
\begin{equation}
\partial_\mu \left(\sqrt{-g}F^{\mu\nu}(-F_{\alpha\beta}F^{\alpha\beta})^{-1/4}\right)=0;
\end{equation}
which, for a constant scalar curvature $R_0$, admit a solution of the form (\ref{metric-sphe}) with the metric function \cite{hendi14a}
\begin{equation}
A(r)=-M-\frac{\left( 2\mathcal{Q}^{2}\right)
^{3/4}}{2\left( 1+f'(R_0)\right) r}-\frac{r^{2}R_{0}}{6},
\label{metric_ads}
\end{equation}
where $M$ is the mass and $\mathcal{Q}$ is the charge. The electromagnetic field has 
\begin{equation}
F_{tr}=\frac{\mathcal{Q}}{r^{2}},
\end{equation}
as the only non-null independent component. We require that $R_0<0$, so the spacetime is asymptotically anti--de Sitter. If $\mathcal{Q}=0$ the well known vacuum static BTZ geometry \cite{btz} in general relativity is recovered, which is also a solution with constant scalar curvature in $F(R)$ gravity \cite{hendi14b}. From the trace of the field equations 
\begin{equation}
R_0 \left(1+f'(R_0)\right)-\frac{3}{2} \left(R_0+f(R_0)\right)=0 
\label{trace1}
\end{equation}
is easy to see that
\begin{equation}
R_0=\frac{3f(R_0)}{2f'(R_0)-1} \equiv 6\Lambda_e ,
\label{trace2}
\end{equation}
in which the effective cosmological constant $\Lambda_e$ is defined. We can introduce an effective charge
\begin{equation}
Z=\frac{\left( 2\mathcal{Q}^{2}\right)^{3/4}}{2\left( 1+f'(R_0)\right)},
\label{defZ}
\end{equation}
which can be positive or negative, having the same sign as $F'(R_0)= 1+f'(R_0)$. The effective Newton constant $G_{\mathrm{eff}} = G/F'(R) = 1/F'(R)$ is positive when $F'(R)>0$, preventing the graviton to be a ghost \cite{revfr}; for a further discussion see Ref. \cite{bronnikov}. Then, $Z>0$ avoids the presence of ghosts, while $Z<0$ requires them. Note that for a particular choice of both the $F(R)$ theory and of the curvature scalar $R_0$, the sign of $Z$ is fixed, since the squared charge $\mathcal{Q}^{2}$ can modify the absolute value of $Z$ but not its sign. The geometry is singular at $r=0$ because the Kretschmann scalar diverges \cite{hendi14a}. The radii of the horizons are determined by the real and positive solutions of the equation $A(r)=0$, which for $Z\neq0$ is equivalent to a cubic equation\footnote{The corresponding analytic expressions are cumbersome, so they are not shown here.\label{cumbersome}} and for $Z=0$ to a quadratic equation. If $Z\ge 0$, the only solution with radius $r_h$ corresponds to the event horizon; in particular, $r_h=\sqrt{-6M/R_0}$ for $Z=0$. When $Z_c<Z<0$, with $Z_c= (-2M/3)\sqrt{-2M/R_0}$, there are two solutions, one corresponding to the event horizon with radius $r_h$, and the other to an inner horizon with radius  $r_i < r_h$; they fuse into one when $Z=Z_c$. If $Z < Z_c$ no horizons are present and the singularity is naked.

In what follows, we adopt the metric function (\ref{metric_ads}) with the intention to provide concrete examples of the formalism introduced in Sec. \ref{tsh}. In them, we construct static thin shells with radius $a_0$ and we analyze their stability under radial perturbations. It is of interest to consider the weak energy condition (WEC), which in the orthonormal basis takes the form of the inequalities $\sigma_0 \geq 0$ and $\sigma_0 + p_0 \geq 0$, in order to determine the type of matter at the shell, being normal when satisfied and exotic otherwise.

\subsection{Charged bubble}

\begin{figure}[t!]
\centering
\includegraphics[width=0.9\textwidth]{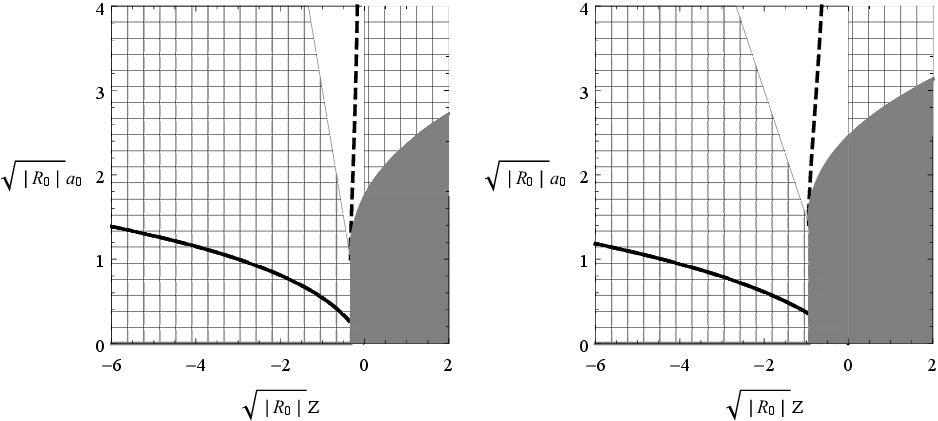}
\caption{Circular bubble with the same scalar curvature $R_0$ for the inner and the outer regions, in a general $F(R)$ theory. The vacuum region is surrounded by a charged thin shell, with radius $a_0$, joining it with a region with mass $M$ and charge $\mathcal{Q}$. The parameter $Z=\left( 2\mathcal{Q}^{2}\right)^{3/4}/\left(2\left( 1+f'(R_0)\right)\right) $ has the same sign as $F'(R_0)= 1+f'(R_0)$ (see text). The solid lines represent the stable solutions while the dotted lines the unstable ones. The meshed zones correspond to normal matter and the gray ones have no physical meaning (see text). Left: $M=0.5$; right: $M=1$. }
\label{burb_R_igu}
\end{figure}
\begin{figure}[t!]
\centering
\includegraphics[width=0.9\textwidth]{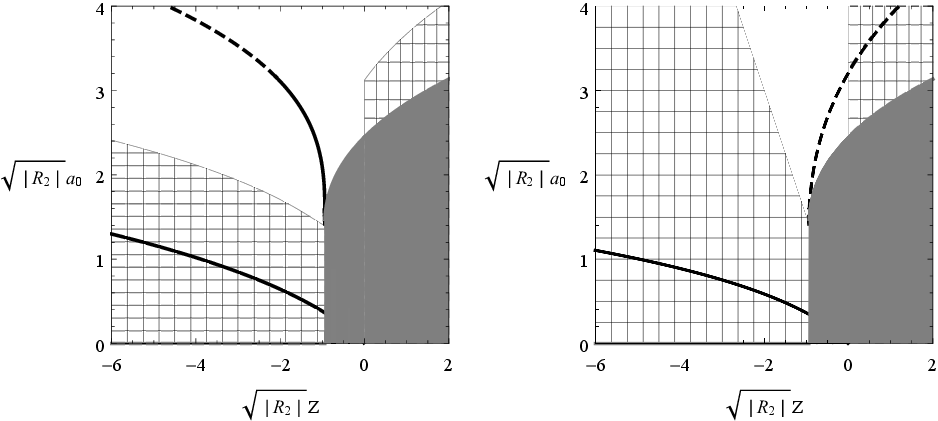}
\caption{Circular bubble with different values $R_1\neq R_2$ of the scalar curvature for the inner and the outer regions, only in quadratic $F(R)$ theory. The vacuum thin shell, with radius $a_0$, is surrounded by a region with mass $M$ and charge $\mathcal{Q}$. The parameter $Z=\left( 2\mathcal{Q}^{2}\right)^{3/4}/\left(2\left( 1+2\alpha R_2\right)\right) $ has the same sign as $F'(R_2)= 1+2\alpha R_2$ (see text). The solid and dotted lines, and the meshed and gray zones have the same meaning as in Fig. \ref{burb_R_igu}. In both plots: $M=1$, $\alpha R_2=-2/5$ for $Z>0$ and $\alpha R_2 =-3/5$ for $Z<0$; left: $R_1 =0.8 R_2$; right: $R_1 =1.2 R_2$. }
\label{burb_R_dif}
\end{figure} 

We first construct a static circular bubble, so the manifold $\mathcal{M}$ has an inner vacuum region $\mathcal{M}_1$ that is joined to an outer region $\mathcal{M}_2$ by a thin shell of matter at $\Sigma$, with radius $a_0$. The interior metric is given by Eq. (\ref{metric_ads}) with constant negative scalar curvature $R_1$ and both null mass and charge, while the outer one by the same equation with constant negative scalar curvature $R_2$, mass $M$ and charge $\mathcal{Q}$. When $Z \ge Z_c = (-2M/3)\sqrt{-2M/R_2}$, the value of $a_0$ is taken larger than $r_h^{(2)}$ in order to remove the region inside the horizon corresponding to the original geometry $\mathcal{M}_2$. The proper matching at $\Sigma$ requires that $a_0$ can only take values satisfying Eq. (\ref{cond-stat}). A positive sign of the second derivative of the potential, which is obtained by replacing the metric functions and their derivatives in Eq. (\ref{potd2}), determines that a configuration with radius $a_0$ is stable under radial perturbations. As it has been previously explained, we have two possible cases to consider:
\begin{itemize}
\item The scalar curvature has the same constant value at both sides of the shell $\Sigma $, i.e. $R_1=R_2=R_0$, and we work in an arbitrary $F(R)$ theory. We use the above defined parameter $Z=\left( 2\mathcal{Q}^{2}\right)^{3/4}/\left(2\left( 1+f'(R_0)\right)\right) $, which has the same sign as $F'(R_0)= 1+f'(R_0)$; this sign is fixed in a particular theory once the value of $R_0$ has been selected. The energy density $\sigma_0$ and the pressure $p_0$ are obtained by replacing the metric functions in Eqs. (\ref{energy-stat1}) and (\ref{press-stat1}), respectively. Some representative results are shown in Fig. \ref{burb_R_igu}. In the left plot we have adopted $M=0.5$ and in the right one $M=1$.
\item The values of the scalar curvature are constant but different at the sides of the shell, i.e. $R_1\neq R_2$, therefore we can only work in the quadratic theory $F(R)= R+\alpha R^2-2\Lambda$, with $\Lambda$ the cosmological constant. The values of  $\Lambda_e$ and  $\Lambda$ are related by Eq. (\ref{trace2}) and they are generally different. Now we define the parameter $Z=\left( 2\mathcal{Q}^{2}\right)^{3/4}/\left(2\left( 1+2\alpha R_2\right)\right) $, which has the same sign as $F'(R_2)= 1+2\alpha R_2$. The energy density $\sigma_0$, the pressure $p_0$, and the external scalar pressure or tension $\mathcal{T}_0$ are calculated by replacing the metric functions in Eqs. (\ref{energy-stat3}), (\ref{press-stat2}), and (\ref{extpress-stat}), respectively; while the strength $\mathcal{P}_{\hat{\imath}\hat{\jmath}}^{(0)}$ of the double layer energy-momentum distribution $\mathcal{T}_{\mu \nu }^{(0)}$ is shown in Eq. (\ref{strength-dyna}). Some representative results are displayed in Fig. \ref{burb_R_dif}. In both plots we have taken $M=1$, $\alpha R_2=-2/5$ for $Z>0$ and $\alpha R_2 =-3/5$ for $Z<0$; while the relation between the values of the scalar curvature is $R_1 =0.8 R_2$ in the left plot and $R_1 =1.2 R_2$ in the right one.
\end{itemize}
In all plots, the solid lines correspond to stable static solutions, while the dotted lines to the unstable ones. The meshed zones represent normal matter satisfying the weak energy condition and the gray areas have no physical meaning, corresponding to the removed part of $\mathcal{M}_2$. In the first case, for $Z<Z_c<0$ there is a stable solution constituted of normal matter, while if $Z_c<Z<0$ another unstable one is present and made of exotic matter; for $Z \ge 0$ no solutions are found. In the second case, there are two possibilities. When $|R_1| < |R_2|$, for $Z<Z_c<0$ there exist two solutions, the one with smaller radius is stable and constituted of normal matter, while the other depending on the value of $Z$ can be stable or not and always made of exotic matter; for $Z \ge Z_c$ no solutions are found. When $|R_1| > |R_2|$, for $Z<Z_c<0$ there is one stable solution constituted of normal matter; for $Z_c<Z<0$ there is one unstable solution with exotic matter; if $Z = 0$ one unstable solution is found with normal or exotic matter (depending also on the value of $\alpha$); for $Z>0$ there is one unstable solution constituted of normal matter. In both cases, changing the value of the mass $M$ associated to the shell does not affect the qualitative behavior of the solutions, it only modifies the scale.

\subsection{Charged thin shell surrounding a black hole} 

\begin{figure}[t!]
\centering
\includegraphics[width=0.9\textwidth]{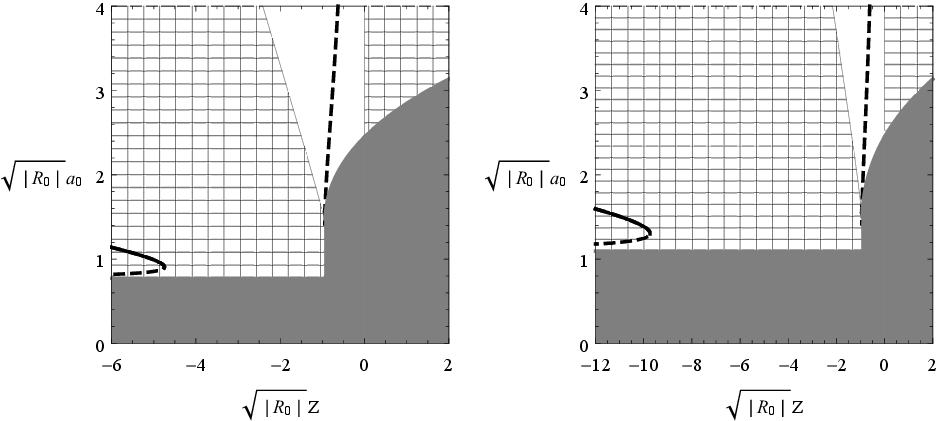}
\caption{Circular shell surrounding a black hole with the same scalar curvature $R_0$ for the inner and the outer regions, in a general $F(R)$ theory. The non-charged black hole with mass $M_1$ is surrounded by a thin shell, with radius $a_0$, connecting to a region with values of mass $M_2$ and charge $\mathcal{Q}$. The parameter $Z=\left( 2\mathcal{Q}^{2}\right)^{3/4}/\left(2\left( 1+f'(R_0)\right)\right) $ has the same sign as $F'(R_0)= 1+f'(R_0)$ (see text). The solid and dotted lines, and the meshed and gray zones, have the same meaning as in Fig. \ref{burb_R_igu}.  Left: $M_1=0.1$ and $M_2=1$; right: $M_1=0.2$ and $M_2=1$ (note in this case the different scale of the horizontal axis). }
\label{casc_BH_Rigu}
\end{figure}
\begin{figure}[t!]
\centering
\includegraphics[width=0.9\textwidth]{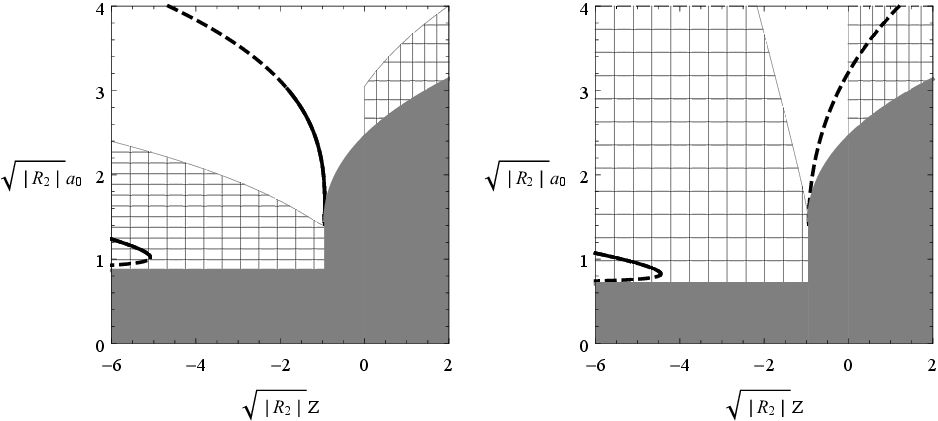}
\caption{Circular shell surrounding a black hole with different values $R_1\neq R_2$ of the scalar curvature for the inner and the outer regions, only in quadratic $F(R)$ theory. The non-charged black hole with mass $M_1$ is surrounded by a thin shell, with radius $a_0$, connecting to a region with mass $M_2$ and charge $\mathcal{Q}$. The parameter $Z=\left( 2\mathcal{Q}^{2}\right)^{3/4}/\left(2\left( 1+2\alpha R_2\right)\right) $ has the same sign as $F'(R_2)= 1+2\alpha R_2$ (see text). The solid and dotted lines, and the meshed and gray zones, have the same meaning as in Fig. \ref{burb_R_igu}. In both plots: $M_1=0.1$ and $M_2=1$, $\alpha R_2=-2/5$ for $Z>0$ and $\alpha R_2 =-3/5$ for $Z<0$; left: $R_1 =0.8 R_2$; right: $R_1 =1.2 R_2$. }
\label{casc_BH_Rdif}
\end{figure}

As a second example, we present a circular charged thin shell surrounding a non-charged black hole. Then, the manifold $\mathcal{M}$ has an inner region $\mathcal{M}_1$ corresponding to a black hole that is joined to an outer region $\mathcal{M}_2$ by a thin shell of matter $\Sigma$, with radius $a_0$. The interior metric is given by Eq. (\ref{metric_ads}) with constant negative  scalar curvature $R_1$, mass $M_1$ and null charge, while the outer one by the same equation with constant  negative  scalar curvature $R_2$, mass $M_2$ and charge $\mathcal{Q}$. When $Z \ge Z_c = (-2M/3)\sqrt{-2M/R_2}$, we adopt $a_0>r_h^{(2)}$ to eliminate the region inside the horizon of the original geometry from $\mathcal{M}_2$. Also, the radius $a_0$ is taken larger than the radius of the event horizon $r_h=r_h^{(1)}=\sqrt{-6M_1/R_1}$ of the black hole. Again, we have two possible cases to analyze:
\begin{itemize}
\item The scalar curvature has the same value at both sides of $\Sigma$, i.e. $R_1=R_2=R_0$, we work in any $F(R)$ theory. We use again the parameter $Z=\left( 2\mathcal{Q}^{2}\right)^{3/4}/\left(2\left( 1+f'(R_0)\right)\right) $, having the same sign as $F'(R_0)= 1+f'(R_0)$, which is fixed once the particular theory and the value of $R_0$ are both selected. The energy density $\sigma_0$ and the pressure $p_0$ are obtained by replacing the metric functions in Eqs. (\ref{energy-stat1}) and (\ref{press-stat1}), respectively. Some representative results are shown in Fig. \ref{casc_BH_Rigu}. In the left plot we have taken $M_1=0.1$ for the black hole mass and $M_2=1$, while in the right one the values are $M_1=0.2$ and $M_2=1$
\item The values of the scalar curvature are constant but different at the sides of the shell, i.e. $R_1\neq R_2$, so we can only work in the quadratic theory $F(R)= R+\alpha R^2-2\Lambda$, with $\Lambda$ the cosmological constant. As previously mentioned, the values of  $\Lambda_e$ and  $\Lambda$ are related by Eq. (\ref{trace2}) and they are generally different. The parameter $Z=\left( 2\mathcal{Q}^{2}\right)^{3/4}/\left(2\left( 1+2\alpha R_2\right)\right) $ has the same sign as $F'(R_2)= 1+2\alpha R_2$. The energy density $\sigma_0$, the pressure $p_0$, and the external scalar pressure or tension $\mathcal{T}_0$ are obtained by replacing the metric functions in Eqs. (\ref{energy-stat3}), (\ref{press-stat2}), and (\ref{extpress-stat}), respectively; while the strength $\mathcal{P}_{\hat{\imath}\hat{\jmath}}^{(0)}$ of the double layer energy-momentum distribution $\mathcal{T}_{\mu \nu }^{(0)}$ is shown in Eq. (\ref{strength-dyna}). Some representative results are displayed in Fig. \ref{casc_BH_Rdif}. In both plots we have adopted $M_1=0.1$ and $M_2=1$ for the masses, $\alpha R_2=-2/5$ for $Z>0$ and $\alpha R_2 =-3/5$ for $Z<0$; while the relation between the values of the scalar curvature is $R_1 =0.8 R_2$ in the left plot and $R_1 =1.2 R_2$ in the right one.
\end{itemize}
The meaning of the solid and dotted lines, and of the meshed zones, is the same as above in all plots. The gray areas have no physical meaning, now corresponding to the removed part of $\mathcal{M}_2$ or the region inside the event horizon of the black hole in $\mathcal{M}_1$. In the first case, within the range $Z<Z_c<0$ there is a pair of solutions, both constituted of normal matter, the one with larger radius is stable and the other is unstable; another unstable solution made of exotic matter also exists when $Z_c<Z<0$; for $Z \ge 0$ no solutions are found. In the second case, there are two possibilities. When $|R_1| < |R_2|$, within the range $Z<Z_c<0$ there is a pair of solutions both constituted of normal matter, the one with larger radius is stable and the other unstable, besides them a third solution made of exotic matter, that depending on the value of $Z$ can be stable or not; for $Z \ge Z_c$ no solutions are found. When $|R_1| > |R_2|$, within the range $Z<Z_c<0$ there is a pair of solutions both constituted of normal matter, the one with larger radius is stable and the other unstable, while another unstable solution with exotic matter is present when $Z_c<Z<0$; for $Z = 0$ one unstable solution is found with normal or exotic matter (depending also on the value of $\alpha$); for $Z>0$ there is one unstable solution constituted of normal matter. In both cases, changing the values of the masses $M_1$ and$M_2$ does not affect the qualitative behavior of the solutions, it only modifies the scale.

\section{Conclusions}\label{conclu}

In this article, we have studied a broad family of (2+1)-dimensional spacetimes with a thin shell of matter, within the framework of $F(R)$ theories of gravity with constant scalar curvature $R$. In our construction, a manifold $\mathcal{M}$ has a circular thin shell $\Sigma$ of matter that joins an inner region $\mathcal{M}_1$ with an outer one $\mathcal{M}_2$. We have analyzed the matter content at $\Sigma$ and the stability of the static configurations, with radius $a_0$, under perturbations preserving the symmetry.

In order to exemplify, we have constructed circular bubbles and thin shells of matter around black holes, within $F(R)$ theories of gravity coupled to conformally invariant non-linear electrodynamics. In the first example, the charged bubble encloses a vacuum region, while in the second one the charged thin shell surrounds a non-charged black hole; in both cases, the spacetime is asymptotically anti--de Sitter. We have allowed the scalar curvature to take the same or different values at the sides of the shell, but in the last case we have been restricted to the quadratic  $F(R)$ gravity, as required by the junction conditions. We have studied the matter content of the shell having a squared charge $\mathcal{Q}^2$ and the stability of the static configurations under radial perturbations. In all scenarios, we have found that stable solutions made of normal matter --with the energy density $\sigma_0$ and the pressure $p_0$ satisfying WEC, i.e. $\sigma_0 \ge 0$ and $\sigma_0 +p_0 \ge 0$-- are possible for suitable values of the parameters, but require ghost fields. When the curvature scalar takes the same value $R_0$ at both sides of $\Sigma $, the existence of solutions --stable or not-- demands that $F'(R_0)<0$, which means that ghost fields are present at both regions of $\mathcal{M}$. When the values of the curvature scalar are different, if $|R_1| < |R_2|$, the presence of the solutions --stable or not--  requires that $F'(R_2)<0$, so they can be found only in the presence of ghost fields in the outer region. Solutions constituted of normal matter without ghosts in the outer region,  i.e. $F'(R_2)>0$, only exist for the case with $|R_1| > |R_2|$, but they are unstable. In the particular case of non-charged shells, i.e. $\mathcal{Q}^2=0$, constructed by using the vacuum BTZ geometry, there exist solutions only when $|R_1| > |R_2|$, which are unstable and can have normal or exotic matter depending on the value of the quadratic coefficient $\alpha$ of the theory; the presence or not of ghost fields also depends on $\alpha$. The main difference in behavior between the two examples is that the bubble requires smaller values of $\mathcal{Q}^2$ --with the same values of the other parameters-- for obtaining the stable solution with normal matter than  the shell surrounding a black hole. In both examples, two non-null extra contributions are also present in quadratic $F(R)$, consisting of the external scalar pressure or tension $\mathcal{T}_0 = \sigma_0 - p_0$ and the double layer energy-momentum distribution $\mathcal{T}_{\mu \nu }$  with a strength $\mathcal{P}_{\hat{\imath}\hat{\jmath}}$ proportional to $\alpha (R_2 - R_1 )$.

\section*{Acknowledgments}

This work has been supported by CONICET and Universidad de Buenos Aires.

\end{document}